\def\year{2024}\relax
\documentclass[letterpaper]{article}
\usepackage{aaai24}  
\usepackage{times} 
\usepackage{helvet} 
\usepackage{courier} 
\usepackage[hyphens]{url} 
\usepackage{graphicx} 
\usepackage{graphicx}  
\usepackage{natbib}  
\usepackage{caption}  
\DeclareCaptionStyle{ruled}%
  {labelfont=normalfont,labelsep=colon,strut=off}
\usepackage{algorithmicx}
\usepackage{algorithm}
\usepackage[noend]{algcompatible}
\newcommand{\algorithmicinput}{\textbf{input}}
\newcommand{\INPUT}{\item[\algorithmicinput]}
\def\BibTeX{{\rm B\kern-.05em{\sc i\kern-.025em b}\kern-.08em
    T\kern-.1667em\lower.7ex\hbox{E}\kern-.125emX}}
    
\usepackage{newfloat}
\usepackage{listings}
\usepackage{amsmath}
\usepackage{appendix}
\DeclareCaptionStyle{ruled}{labelfont=normalfont,labelsep=colon,strut=off} 
\floatstyle{ruled}
\newfloat{listing}{tb}{lst}{}
\floatname{listing}{Listing}
\usepackage[table,x11names]{xcolor}

\newcommand{\dwbd}{$\text{DW-BD}$}

\title{Bridging Social Media and Search Engines: Dredge Words and the Detection of Unreliable Domains}
\author {
    Evan M. Williams\textsuperscript{\rm 1},
    Peter Carragher\textsuperscript{\rm 1},
    Kathleen M. Carley\textsuperscript{\rm 1}
}
\affiliations {
    \textsuperscript{\rm 1}Carnegie Mellon University\\
    \{emwillia, pcarragh, carley\}@andrew.cmu.edu
}

\usepackage{bibentry}

\begin{document}

\maketitle

\begin{abstract}

Proactive content moderation requires platforms to rapidly and continuously evaluate the credibility of websites. Leveraging the direct and indirect paths users follow to unreliable websites, we develop a website credibility classification and discovery system that integrates both webgraph and large-scale social media contexts. We additionally introduce the concept of \textit{dredge words}---terms or phrases for which unreliable domains rank highly on search engines---and provide the first exploration of their usage on social media. Our graph neural networks that combine webgraph and social media contexts generate to state-of-the-art results in website credibility classification and significantly improves the top-k identification of unreliable domains. Additionally, we release a novel dataset of dredge words, highlighting their strong connections to both social media and online commerce platforms.

\end{abstract}

\section{Introduction}

On February 24, 2022, the day Russia began its invasion of Ukraine, Jacob Creech, a fringe QAnon conspiracy theorist, posted a series of unsubstantiated claims on Twitter. These tweets insinuated that the U.S. had created COVID-19 and implied that Russia’s invasion was actually an effort to shut down U.S.-funded ``biolabs'' in Ukraine and prevent another global pandemic \cite{adl2022}. Hours after these tweets, the conspiratorial website InfoWars published an article promoting them, crediting Creech for uncovering an "ulterior motive theory" \cite{adl2022}. The tweets soon circulated to other conspiratorial sites, and search interest in ``U.S. biolabs'' and ``Ukraine biolabs'' spiked in the following days \footnote{\url{https://trends.google.com/trends/explore?date=today\%205-y&geo=US&q=Ukraine\%20biolabs&hl=en}}. Conspiratorial sites were likely the predominant search results until Snopes debunked the claim later that day \cite{snopes2022}.


In some ways, this represents a success of the current misinformation response paradigm---fact-checkers acted swiftly to address a rapidly spreading falsehood. However, it also highlights the limitations of reactive misinformation interventions. Despite the swift debunking by fact-checkers, the conspiracy theory was still echoed by established news sources like Fox News and later amplified on the floor of the U.S. Senate \cite{adl2022}. Researchers later found that a coordinated network of social media accounts helped boost the narrative on Twitter \cite{alieva2022investigating}. As of 2024, even though fact-checkers have repeatedly debunked the underlying claims \cite{nprbiolab2022,rand2022}, a search engine audit study found that Google, Bing, and Yandex still return first-page results that promote the ``Ukraine biolabs'' conspiracy theory \cite{kuznetsova2024algorithmically}. This example highlights an important and understudied aspect of misinformation: the interaction between social media and search engines.

An alternative to reactive fact-checking is proactive algorithmic content moderation. This can involve modifying recommendation and ranking systems to reduce the reach and virality of unreliable information sources. In search engines, this might mean downranking articles from unreliable domains, while on social media, it could mean ranking posts containing unreliable links lower in users' newsfeeds. For proactive approaches to work, platforms need systems that can identify unreliable domains. Both classification and discovery are essential as unreliable websites can and do employ tactics to evade blacklists \cite{carragher2024detection}. In this work, we present a novel approach to unreliable domain detection and discovery that leverages signals from both large-scale social media data and webgraph data.

Little is understood about how users transition from social media to search engines to access unreliable websites. However, it is known that directing users to search engines can be an effective tactic for spreading misinformation. Since 2016, global surveys have consistently found that individuals trust search engines more than traditional media \cite{barometer2024edelman,qz2015google}. Thus by directing users to search engines, either explicitly or implicitly, bad actors can foster a false sense of content reliability. Research has only recently begun exploring these pathways, and some helpful concepts like \textit{problematic queries} \cite{golebiewski2019data}, \textit{data voids} \cite{golebiewski2019data}, and \textit{keyword signaling} \cite{tripodi2019}, have been examined in case studies. The recently proposed concept of \textit{search directives} offers a clearer and more observable path by which users are directed to unreliable content. A search directive is content explicitly intended to prompt an online search, such as user $a$ telling user $b$ to "look up Chemtrails on Google" \cite{robertson2023identifying}.

However, search directives have two key limitations as content moderation tools. First, while they can lead users to unreliable content, they can also direct them to neutral or beneficial information, such as song lyrics or mathematical theorems \cite{robertson2025data}. Second, as seen in the ``Ukraine biolabs'' case, users can be driven to unreliable content through search engines even without being explicitly told to search something. To overcome these limitations, we propose the concept of \textit{dredge words}---terms or keyphrases for which unreliable domains rank highly in search results.

By attempting to explicitly incorporate each of the paths that users take to unreliable websites into GNNs, we seek to explore the dynamics that connect the spread of unreliable content on social media and search engines. We demonstrate that our relatively-simple curriculum-based heterogeneous graph model that leverage context from both webgraphs and social media data achieves SoTA results on the website credibility classification task. Further, our best model model does not incorporate any explicit text or semantic content from the webpages or from social media users, which makes this approach flexible, and easily-extendable to non-English contexts.

Finally, we provide the first exploration of \textit{dredge words} on social media. We incorporate a small set of dredge words into an unreliable domain discovery process in an attempt to mimic how social media users may transition from social platforms to the discovery of misinformation sources via search engines. Surprisingly, we find that dredge words frequently surface social media URLs in top Google Search Engine Result Page (SERP) positions---i.e., the websites returned when a user enters a query on Google Search. This suggests the existence of a bidirectional path that often leads back to social media. We show our heterogeneous model greatly outperforms competing systems in the top-k discovery of unlabeled unreliable websites. We show our heterogeneous model greatly outperforms competing systems in the top-k discovery of unlabeled unreliable websites. We publicly release the code, webgraph data collected for this project, a dataset of 3,939 dredge words for 46 unreliable news domains, and the resulting SERPs for each of the dredge word  queries\footnote{\url{https://github.com/CASOS-IDeaS-CMU/DredgeWords}}. Our findings demonstrate that both direct and indirect paths to misinformation provide important signals that can be leveraged by researchers and platform designers working to mitigate the spread of misinformation across digital ecosystems.

\section{Related Works}

\subsection{Social Media and Webgraphs}
Classifying individual texts or articles is a core problem addressed by research in misinformation detection. Such detection methods have typically relied on website content and social media data \cite{page_detection_topic_agnostic,domain_discovery_social_media,domain_discovery_sm_page_detection_sm_content,wang2024news}. Castelo et al. proposed a topic-agnostic detection system \cite{page_detection_topic_agnostic} that identifies unreliable articles based on Linguistic Inquiry and Word Count features. Chen and Freire adopted the method for the task of unreliable domain discovery \cite{domain_discovery_social_media}. Their discovery system capitalized on user tendencies within the social graph, wherein a user who tweets a URL from a known unreliable source is likely to also tweet URLs from yet unknown sources. Similarly, Silva et al. combine website content and social context resulting in a misinformation page detection method that leverages heterogeneous data types \cite{domain_discovery_sm_page_detection_sm_content}, with a focus on early detection across a broad range of topics. The predictive power of webgraphs and social media data have been demonstrated separately in several contexts. Aswani et al. detect SEO manipulation by link-building sites by clustering on Pagerank score and Domain Authority \cite{link_scheme_detection}. For detecting news site bias, Aires et al. \citet{www_bias_detection} scrape cross-links from a list of prominent news sites from Media Bias Fact Check \citep{mbfc}. Sehgal et al. explore a case where misinformation was spread through coordinated hyperlink and social media networks \cite{link_scheme_misinfo}. Another recent case study demonstrated the manipulated webgraph linkages of unreliable pseudo-thinktanks \cite{williams2023search}. Additionally, Zhang and Cabage show that social media and SEO promotion have different strengths; they find that social-sharing results in immediate but short-term boosts to traffic, while the benefits of link-building are slower to realize, but last longer \cite{social_media_vs_link_building}. This hints that a combination of both webgraph and social media data should be used in investigating misinformation sources. However, to our knowledge, this is the first work that attempts to combine the two contexts for unreliable domain classification or discovery. 

Despite its importance and relevance, domain-level credibility detection is a relatively uncommon task in the literature---partially as a result of the lack of accepted labels. The most comprehensive label list as of 2024 are those published in \citet{lin2023high}, which aggregates the scores of multiple different domain reliability rating lists. The vast majority of work in this space occurs at the article level (e.g., \citet{nakov2023overview,bianchi2024evaluating}). This class of approaches often take a sample of articles from each website and aggregate reliability ratings in some way. These approaches have several important limitations; namely, article-level approaches are restricted to languages with LLM support, model qualities depend heavily on what is sampled, many unreliable news sites often report true information with heavily partisan slants, and articles don't necessarily reveal the goals of the publisher---e.g., knowing a media outlet has strong ties with a adversarial government can change how readers interpret the site's content. Consequently, researchers have recently begun exploring content-agnostic domain classification approaches. In 2023, \citet{yang2023large} showed that LLMs could be used to evaluate website credibility. More recently, \citet{carragher2024detection} proposed a webgraph-based model for unreliable domain detection and discovery tasks. For discovery, the authors use backlinking domains in the webgraph in a snowball sampling approach. The authors demonstrate that GNN models trained on SEO attributes and webgraph data are effective for detection.

\section{Data}

\begin{figure}[!ht]
    \centering
    \includegraphics[height=5.2cm]{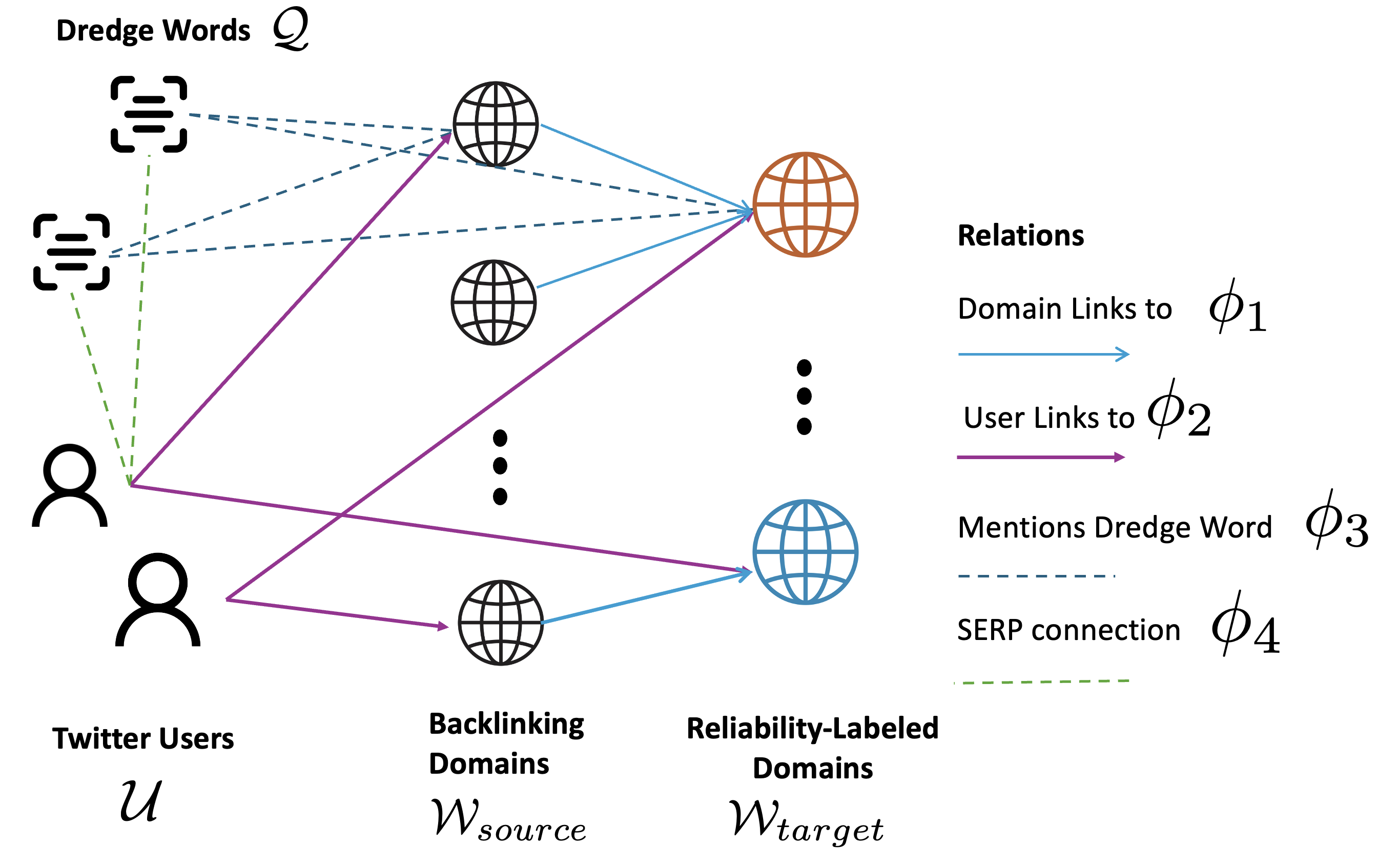}
    \caption{A summary of the heterogeneous graph construction process. Solid lines denote direct paths (a user clicks a hyperlink), and dashed lines denote indirect paths (a user sees a post and then queries a subset of that post on a search engine).}
    \label{fig:data_pipleine}
\end{figure}

We construct a heterogeneous graph with different relations based on the direct and indirect paths that users can take to unreliable websites. To capture direct paths to target news sites, we collect the 10 domains which most frequently link to each labeled news site (e.g., website $i$ has linked 1M times to website $j$). We further extract social media mentions of each target news site (e.g., user $k$ posts a hyperlink to website $j$). We consider both of these to be \textit{direct paths}, as in both these cases, users would simply click a link to access website $j$. Finally we consider an \textit{indirect path} through dredge words. Dredge words connect users to websites through Google SERPs. When browsing social media, users can encounter words or concepts with which they're unfamiliar and that exist in data voids---for example, a user might see a Twitter post about ``indigo children'' and search a variation of that query on Google. The results surface numerous pseudoscientific websites and advertisements (see Appendix A). We provide a summary visualization of this heterogeneous network in Figure \ref{fig:data_pipleine}.

\subsection{Webgraph and Features}

We use the domain credibility labels published by \citet{lin2023high}. The authors of the work align the domain reliability ratings of all domains ranked by 6 expert groups and run imputation followed by principal component analysis to generate aggregate ratings for 11,520 domains. While the the sites are largely English-language, there are websites spanning multiple languages and target audiences, including news sites that operate in Italian, Russian, and German. We binarize the news domain rankings as "reliable" and "unreliable" using a threshhold of 0.5162, corresponding to the bottom two quintiles of the data. Following the methodology of \citet{carragher2024detection}, We use the SEO toolkit service Ahrefs\footnote{\url{ahrefs.com}} to extract the 10 domains which link to each of the 11,520 target domains at the highest volume (the highest-volume back-linking domains). Ahrefs feature values have been found to correlate strongly with ground-truth pagerank calculations and competing traffic estimate systems  \cite{carraghermisinformation}. 
In total, we extracted 43,758 domains, each with 23 domain-level attributes. Additional discussion of labels, collection details, and EDA of the webgraph can be found in Appendix A. Each edge in this dataset denotes a direct path, as a user on the source site could click a hyperlink to move to the reliability-labeled target site. We were unable to extract features for 193 domains---the domains largely appeared dead or inactive, so we excluded them from the data. This left us with 11,327 labeled domains.


\subsection{Twitter Data}
For social media context, we use a Twitter dataset constructed by querying COVID-related keywords\footnote{coronavirus, Wuhan virus, Wuhanvirus, 2019nCoV, NCoV, NCoV2019, covid-19, covid19, covid 19} via Twitter's streaming API between January 29, 2020 and June 26, 2022. Due to server issues and API limitations, 121 days over the time period have partial or missing data. However, these gaps are spread relatively evenly over the time period, and so the data still provide strong coverage. Our final Twitter dataset contained 3.6 billion extracted tweets. 

From this set of 3.6 billion tweets, we extract all Tweets that either link to one of our 11,327 websites or retweet, reply to, or quote a tweet containing one of our 11,327 websites. This resulted in a dataset of 320M tweets that link to over 840k unique domains. Of the original 11,327 labeled domains, only 5,504 appeared within the Twitter data. To reduce the noise in the dataset we perform several cleaning and filtration operations including dropping all tweets that do not directly link to one of our target sites. Due to length constraints, we detail Twitter cleaning in Appendix A. Following these cleaning operations, we are left with 555k users and 4.9M unique tweets which we observed tweeted or retweeted 91.2M times. This reduced dataset contains mentions of 2,475 reliability-labeled domains and 714 unlabeled backlinking domains. 

\subsection{Dredge Words}

For the 100 websites with the respective lowest and highest reliability rankings, we extract the top-ranked $k \leq$  1,000 Google keyphrases from Ahrefs, for a total of 34,646 keyphrases. We elected to choose a limited set of websites due to an infrastructure bottleneck: querying dredge words over 3.6B tweets is very slow on our hardware. We then used WebSearcher \cite{robertson2020websearcher} to query each of these keyphrases on Google and extracted the first 10 URLs returned in each SERP. There was no user account and cookies were not stored across queries; scraping was conducted from an IP address in Pittsburgh, Pennsylvania, which could impact search results. We kept all queries for which the target unreliable domain was returned in the top 10 Google search results. As many of the domains for which we collected dredge words did not have any queries for which they ranked in the top 10 on Google, this resulted in a set of 3,939 \textit{dredge words} spanning 46 unreliable domains. We created a separate query to find mentions of these keyphrases in the 3.6B tweet covid dataset. This yielded 5.7 million tweets containing dredge words. Many of the most common mentions were explicit mentions of domain names or organization names (which can be identical to twitter handles---like ``gatewaypundit'', ``infowars'', and ``nvic'' all received over 10,000 mentions. To filter down the data, we use regular expressions to ensure each drudge word begins with a hashtag, starts a tweet, or is preceded by a white space. This retains 213 dredge words in 421k tweets that qualitatively contain less noise. Of these 421k tweets, only 9,788 (2\%) explicitly linked to the unreliable domain associated with the dredge word.

\begin{figure}[!ht]
    \centering
    \includegraphics[height=7.5cm]{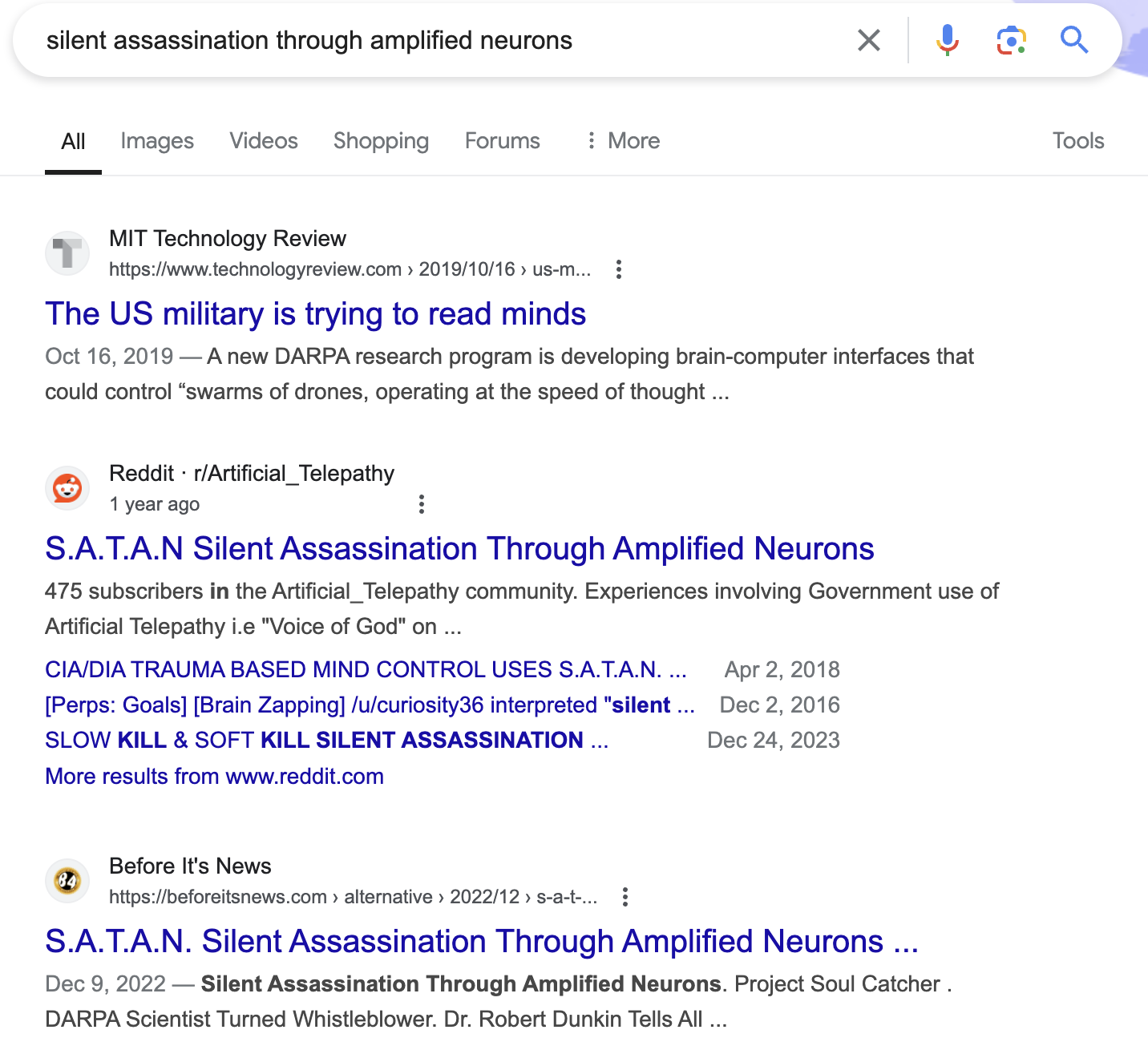}
    \caption{The top search results for the dredge word ``silent assassination through amplified neurons''. The query surfaces fringe reddit subreddits followed by ``beforeitsnews'', an unreliable news source.}
    \label{fig:satan}
\end{figure}

This data pipeline attempts to mirror the way a user might access this information ``in the wild''. A user might encounter a Twitter post that contains a reference to ``silent assassination through amplified neurons'', or ``Project S.A.T.A.N.'', search the query on Google, and encounter unreliable results like those in Figure \ref{fig:satan}. Some additional examples of dredge words include ``psychic attack'' (28 Twitter mentions), ``akashic record'' (12), ``flu shot injury'' (6), and ``fallcabal'' (5). However, we note that our current collection pipeline is almost certainly underestimating their actual usage as 1) on Twitter, we only capture dredge word usage mentioned alongside COVID keywords, and 2) we extracted dredge words in 2024, 2-4 years after tweets were posted. While the majority of the dredge words are in English, other languages are present in the list, including Chinese, Hindi, and Arabic. We use this condensed twitter dataset and the set of SERP results they yielded in our dredge-word-based unreliable domain discovery process. While we were initially interested in paths from social media to search engines, we find dredge word SERPs surprisingly demonstrate a strong paths to social media. Youtube was by far the most commonly-returned domain (4,304 times). Wikipedia, Reddit, Quora, Twitter, Amazon, and Facebook were also among the most commonly-returned domains, in part due to the widespread commercialization of pseudo-scientific concepts.

\section{Case Study}

\begin{figure}[!htbp]
    \centering
    \includegraphics[height=3.6cm]{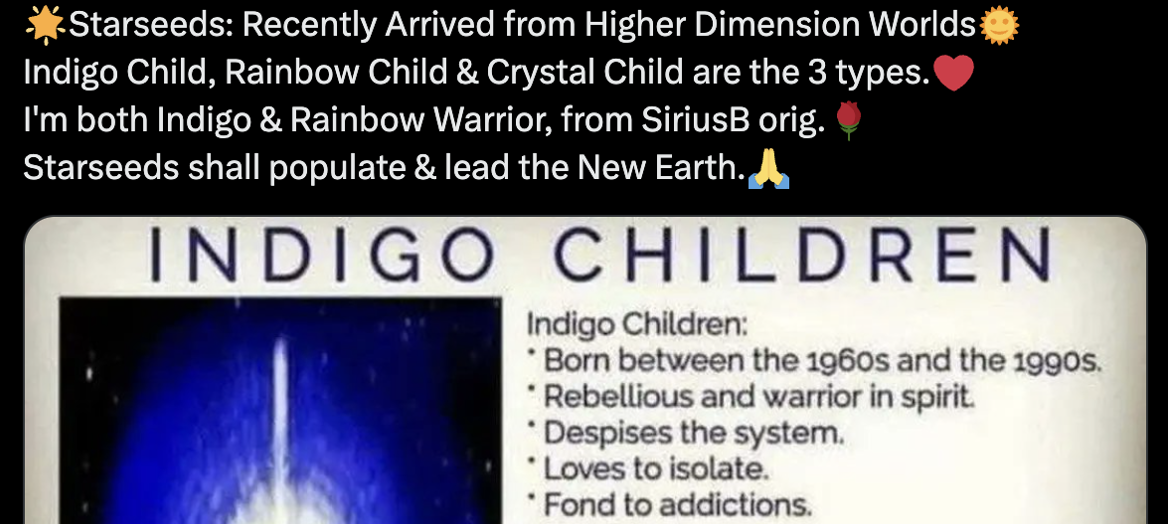}
    \caption{A truncated tweet about ``Indigo Children''.}
    \label{fig:indigo_tweet}
\end{figure}

To provide a more concrete illustration of the tasks and our motivation, we'll consider some of the ways that users might end up on the unreliable website, Gaia\footnote{https://mediabiasfactcheck.com/gaia/}. This website often promotes pseudoscience, vaccine misinformation, and various conspiracy theories. We can imagine two users taking direct paths and a third following an indirect path. The first imaginary user follows paranormal accounts or pages on social media, and sees an account post a link to an article about haunted houses on Gaia. The second individual is interested in UFOs and listens to the ``Coast to Coast AM''\footnote{\url{https://en.wikipedia.org/wiki/Coast_to_Coast_AM}} radio talk show---if that individual visited the website for the talk show, they would find 29.4K hyperlinks that would lead to Gaia. In a third instance, we can imagine a user interested in astrology seeing the post in Figure \ref{fig:indigo_tweet}.

The user, who is encountering new information, might then be curious and compelled to search ``indigo children meaning'' or just ``indigo children'' on Google---both phrases are dredge words that surface Gaia in the top 10 results. Interestingly, after manually searching these ``indigo children meaning'' in June 2024, we found that Google's definition snippet---a summarization box on the top of some SERPs with the text ``From sources across the web''---expresses pseudoscientific concepts. The box states that ``Indigo Children are kids with indigo-colored auras''. This snippet was followed by YouTube videos promoting Indigo Children pseudoscience and a Reddit post from the /r/Psychic subreddit. Another user who may have searched only ``indigo children'' would have seen a snippet from Wikipedia which correctly calls indigo children pseudoscientific. However, adjacent to the snippet, the user would have seen four books for sale about indigo children. After that, the second highest-ranking return was an article from an unreliable domain identifying the ``13 signs you're an indigo child''.


This case study also highlights the importance of conceptualizing misinformation consumption as interconnected; none of our hypothetical users were actively seeking anti-vaccine content, and yet in pursuing their relatively-innocuous interests, all are exposed to medical misinformation. While all of these paths lead to Gaia, the indirect routes may be the most effective at swaying users, as discovering Gaia through a trusted intermediary (in this case, Google) may confer additional trust to the source \cite{robertson2023identifying}. The questions we are trying to answer in this work, with respect to this case study, are 1) ``How does integrating each of these path contexts into a model impact our ability to identify Gaia as unreliable?'' and 2) ``What websites that link to Gaia are also promoting misinformation''? Even in this case study, neither tasks are trivial; detecting unreliable websites is challenging, and while many of the websites that link to Gaia are unreliable, others are generic SEO sites. Additionally, there are reliable and .edu sites that link to Gaia in fact-checking articles, which further complicates the task.
\section{Methods}

In order to evaluate the impact of additional levels of context on our models, we construct graphs that capture increasingly granular levels of context present in the data. Let $\mathcal{W}$ be the set of all labeled and unlabeled websites $\{w_1, w_2, \dots, w_n\}$ contained in the extracted webgraph data. We define $\mathcal{W}_{target}$ as the subset of $\mathcal{W}$ for which we have reliability labels---i.e., the 11,327 reliability-labeled news domains; we define $\mathcal{W}_{source}$ as the 10 websites that most link to each domain. We note that $\mathcal{W}_{source}$ contains some websites in $\mathcal{W}_{target}$. Next define $\mathcal{U}$ as the set of all Twitter users $\{u_1, u_2, \dots, u_n\}$ that link to domains $\in \mathcal{W}$. Finally, let dredge words $\mathcal{Q}$ be the set of keyphrases $\{q_1, q_2, \dots, q_n\}$ connected to SERPs containing URLs $\in \mathcal{W}$.

We define several preliminaries. A Graph Union operation ($G = G_1 \cup G_2$), between graphs $G_1 = (V_1, E_1)$ and $G_2 = (V_2, E_2)$ is defined as $G = (V,E)$ where $V = V_1 \cup V_2$ and $E = E_1 \cup E_2$. A heterogeneous graph is an extension of a homogeneous graph, $\mathcal{G} = (V, E)$, where $V$ and $E$ are associated with a node type mapping function $\Psi : V \rightarrow A$ and an edge type mapping function $\Phi : E \rightarrow \phi$. In our setting, the set of node types are $A = \{\mathcal{W}, \mathcal{U}, \mathcal{Q}\}$ and the set of edge types are $\Phi = \{\phi_1: domain-to-domain, \phi_2: user-to-domain, \phi_3: user-to-dredge word, \phi_4: dredge word-to-domain\}$. This setup is visualized in Figure \ref{fig:data_pipleine}.



\subsection{Homogeneous Graphs}

We consider three homogeneous baselines---where all nodes are treated as part of the same class---to justify the added complexity of using heterogeneous networks. We define two homogeneous graphs: $\mathcal{H}_{domains}$ contains domain-to-domain relationships, i.e., $\mathcal{W}_{source}$ to $\mathcal{W}_{target}$. $\mathcal{H}_{users}$ contains relationships between Twitter users and the websites they mentioned $\in \mathcal{W}$. Finally, we define $\mathcal{H}_{domains + users}$ = $\mathcal{H}_{domains} \cup  
\mathcal{H}_{users}$. Clearly, these latter two graphs contain different sets of nodes and relations, but we treat them as equivalent in these experiments as a simple baseline.

As the features of domains and users are drawn from different spaces and have different sizes, we elect to use positional node features calculated individually on each of the three constructed homogeneous graphs. For this graph, we therefore define $Z_n$ as features calculated using using Node2Vec \cite{grover2016node2vec} parameterized with a walk length to 20, context size to 10, and 10 walks per node, and an embedding dimension of 23, chosen because it is identical to the number of features that we extract for each website from Ahrefs. 

$\mathcal{H}_{users}$ presents an additional challenge in that not all websites $\in \mathcal{W}_{target}$ are mentioned by users $\in \mathcal{U}$. We include social-media-only baselines, but these are trained and evaluated on only a network of user interactions with the 2,475 labeled domains mentioned in the social media data, which make these results not directly comparable with our other models.

\subsection{Additional Baselines}

We include comparisons to two past works that explored website credibility classification without including article or website content. First, we re-implemented the GNN for news domain credibility classification and discovery proposed in \cite{carragher2024detection}, which is a homogeneous model which uses the Ahrefs features of domains. Following the authors, we log-normalize the Ahrefs features of each domain. We train and evaluate their model on all labeled 11,327 reliability-labeled domains. 

Second, we compare our results to those from \cite{yang2023large}, where the authors asked ChatGPT to return the probability that a website is unreliable. Following the authors recommendation, we take a binary cut-off at 0.5 and compare results accordingly. As there is no hold-out set for the ChatGPT data, we consider two comparisons between our best-performing model and this approach. We first compare accuracy and F1 predictions for all 7,318 sites; however, this comparison is problematic as this guarantees leakage---i.e., some of these sites were in our model's training data. To account for leakage, we perform a second evaluation where we only evaluate the predictions of our model and of ChatGPT on the hold-out test set of domains that our model never encountered.

\subsection{Heterogeneous Graphs}

We construct two heterogeneous graphs. $\mathcal{E}_{domains + users}$ is structurally equivalent to its homogeneous counterparts, but explicitly incorporates the two node types ($\mathcal{W}, \mathcal{U}$) and their different relations ($\phi_1, \phi_2$). we use a heterogeneous graph-neural network architecture that can treat user nodes and domain nodes as different nodetypes. We additionally construct $\mathcal{E}_{domains + users + dredge}$, which contains all node types ($\mathcal{W}, \mathcal{U}, \mathcal{Q}$) and all relation types ($\phi_{1:4}$). 

In the heterogeneous networks, the features of domain nodes $\mathcal{W}$ are always the logged domain-level features extracted from Ahrefs, as in \citet{carragher2024detection}. For user nodes, we ran each model with positional context---Node2Vec features, and separately with text context. To create textual context, for each user, we randomly sample, without replacement, 10 tweets for each observed user and embed them using multilingual distilBERT \cite{Sanh2019DistilBERTAD}. This is a naive approach that excludes a large amount of context for some users, particularly as we observed two users tweet over 1 million times in our filtered dataset. However, considering only 10 tweets is a standard practice in user embedding literature \cite{pan2019social}. A clear path for future work is to explore the impact of more advanced user embedding strategies within this system. Dredge word embeddings were extracted using multilingual distilBERT, and these embeddings are used as features for dredge word nodes. We note that dredge words were only considered for 46 of the least reliable websites, so their inclusion in models is likely to harm the models, as those features only exist for a small set of the 11,327 labeled domains. We nonetheless report statistics and discovery evaluations.

\subsection{Graph Neural Network Training}

A label-stratified 80/10/10 split on labeled websites is used to create training, validation, and test sets. Nodes that did not have reliability labels---the majority of nodes in all networks---were masked during training. For our one-mode, heterogeneous, and homogeneous experiments we use a simple 2-layer graph neural network using GraphSAGE convolutions proposed in \cite{hamilton2017inductive}. We ran experiments using graph attention networks \cite{velivckovic2017graph} and heterogeneous graph transformers \cite{hu2020heterogeneous}, but found marginal accuracy gains at the cost of longer training times. As our primary classification interest is evaluating the impact of context---i.e., signals from different sources---on domain credibility classification, rather than finding the best architecture, we elect to use homogeneous and heterogeneous GraphSAGE layers for each respective model. We include a single baseline that uses a heterogeneous graph transformer as well. Each model consists of a SAGEConv layer with dropout, with a hidden dimension of 512, followed by a ReLU activation and a second SAGEConv layer with a log softmax activation function. For models trained with user text, we included an additional linear layer to align input feature sizes. For each model, we train for a maximum of 1,000 epochs with early stopping based on validation loss and a patience of 50. We use an Adam optimizer with a starting learning rate of 1e-3 and a cosine annealing learning rate scheduler\footnote{Implementation is available at \url{https://github.com/CASOS-IDeaS-CMU/DredgeWords}}. The heterogeneous GNN, the most complex network we consider, contains 3.2M parameters, and no model exceeded 5 minutes of training time to reach our early stopping convergence condition. All models in this paper were run or trained on a single NVIDIA GeForce RTX 3080. 



\subsection{Curriculum Learning}

We assume that highly reliable and highly unreliable websites are easier to differentiate in webgraphs than those that are mixed, mostly reliable, or mostly unreliable. A manual examination of the labeled URL data leads us to believe that this assumption is very reasonable when website content is considered (see Appendix B). We implement a slightly-modified version of the``Baby Steps" learning curriculum proposed and explored in \citet{spitkovsky2010baby,cirik2016visualizing}. As domain labels from \citet{lin2023high} contain unified principal component scores of expert ratings of domain reliability, we can develop a curriculum that first learns labels of extremely reliable and extremely unreliable domains, and that gradually works from the extremes towards the websites with mixed reliability. Using the original labeled domain dataset $\mathcal{D}$, we calculate quintiles of reliable labels and unreliable labels using principal component scores, and define these as an ordered set of batches $\{d_1, d_2, \dots, d_{10}\}$, where $d_1$ is the first quintile of the most reliable domains and $d_{10}$ is the fifth quintile of the unreliable domains. When using curriculum learning, we begin training the model $\mathcal{M}$ using $\{d_1, d_{10}\}$, and following convergence, the model considers $\{d_1, d_{10}\} \cup \{d_2, d_9\}$. The model continues in this fashion until it converges on all available data. In our Baby Steps implementation, the we define model's corresponding convergence as 10 epochs without an improvement in validation loss. Once the curriculum has incorporated all data, the convergence criterion described in the previous section are used. More generally, for a website reliability curriculum $\mathcal{C}$, an even-length dataset $\mathcal{D}$, and a model $\mathcal{M}$, our implementation of the Baby Steps curriculum can be expressed as the procedure in Algorithm \ref{alg1}.

\begin{algorithm}[H]
  \caption{Modified Baby Steps Curriculum}
  \begin{algorithmic}
    \INPUT{$\mathcal{M}$, $\mathcal{D}$, $\mathcal{C}$}
    \STATE{sort($\mathcal{D}$, $\mathcal{C}$)}
    \STATE{$B \gets \emptyset$}
    \FOR{$i \in 0, 1,\dots \frac{k}{2}$}
        \WHILE{$\mathcal{M}$ not converged}
            \STATE{$B \gets B \cup \mathcal{D}[i] \cup \mathcal{D}[k-i]$}
            \STATE{$\mathcal{M}(B)$}
        \ENDWHILE
    \ENDFOR
    \STATE{\textbf{return} $\mathcal{M}$}
  \end{algorithmic}
  \label{alg1}
\end{algorithm}

\subsection{Unreliable Domain Discovery}
While we can evaluate our classifiers on domains in $\mathcal{W}_{target}$, we desire to make a tool that can identify unknown unreliable domains. To evaluate our best-performing classifier, we explore its ability to identify unlabeled unreliable domains over the unlabeled domains in $\mathcal{W}_{source}$. We outline our discovery processes and we benchmark our discovery process against two previous works.

We implement and evaluate two distinct discovery processes. The first is GNN discovery, where we take predictions for unlabeled domains in the graph from the best performing GNN model. We call the second method Dredge Word-Based Discovery ($\text{DW-BD}_{lower}$), which mimics the path social media users take to reach unreliable domains by observing dredge words on social media, and querying these terms on a search engine. Using WebSearcher \cite{robertson2020websearcher}, we create a candidate list of domains from the top 10 SERP results for each of the dredge words we have compiled from our Twitter dataset. We then pull Ahrefs attributes for each of the candidate domains and use the SEO attribute classifiers from \citet{carragher2024detection} to filter down the candidate domains as detailed in Appendix A. 

We add an additional Dredge Word-Based Discovery baseline ($\text{DW-BD}_{upper}$), which does not limit the discovery process based on whether or not we observed dredge words mentioned in the Twitter dataset. For this baseline, we extracted Google SERPs for 118k dredge words associted with 1,051 unreliable domains. We then dropped all results where we did not observe the target domain in rank in the first 10 SERP results. This resulted in 73,843 dredge words for which 836 unreliable websites ranked in the top 10 Google results. The resulting dataset contained 884k individual search results. Due to Ahrefs API constraints, we filter randomly drop 30k domains that only appeared once in search results, leaving 854k individual search results linking to 44,503 unique domains. We and run the discovery process described on this dataset above to generate candidate unreliable domains. Whereas $\text{DW-BD}_{lower}$ is bounded by Twitter data, $\text{DW-BD}_{upper}$ is not.

We compare our discovery processes with the webgraph-based discovery (WG-BD) process proposed in \citet{carragher2024detection} and with the social media-based discovery process (SM-BD) proposed in \citet{domain_discovery_social_media}. Specifically, we consider Precision@5, @10, @20, and we consider the partial F1 metric proposed in \citet{domain_discovery_social_media}. This means we run the discovery process twice, once on the full domain list where we evaluate results manually with top-10 and top-20 accuracies, and again on a restricted domain list to compute the partial F1 metric. We additionally report partial precision and recall.

Due to a lack of ground truth labels for evaluating newly discovered domains, Partial precision, recall, and F1 measure the ability of the discovery system to find unreliable domains with respect to two known lists of unreliable sources, a seed list and an evaluation list. As defined by \cite{domain_discovery_social_media}, the seed list is the PoliticalNews dataset and the evaluation list is drawn from unreliable MBFC domains:

\begin{equation}
     {p = \frac{\# \text{prediction = fake \textbf{and} MBFC label = fake}}{\# \text{prediction = fake}}}
\end{equation}

\begin{equation}
    {r = \frac{\# \text{prediction = fake \textbf{and} MBFC label = fake}}{\# \text{MBFC label = fake}}}
\end{equation}

\begin{equation}
    \label{eq:partial_f1}
    {pf1 = 2 \times \frac{p \times r}{p + r}}
\end{equation}

Partial F1, as given by Equation \ref{eq:partial_f1}, measures the ability of a discovery system seeded on PoliticalNews to discover as many MBFC domains as possible (high partial recall), without discovering domains that are not in the MBFC list (high partial precision) \citep{page_detection_topic_agnostic}. These metrics are further limited in our DW-BD experiments, as many (53 of 74) politifact websites are either inactive or do not have keywords that rank them on Google. For those experiments, we therefore calculate partial metrics only on SERPs corresponding to dredge words from 21 active domains. We discuss the limitations of the partial F1 metric in more detail in Limitations and Appendix C.

\section{Results}
\subsection{Credibility Classification}
We present Accuracy and F1 statistics for website reliability classification for homogeneous models ($\mathcal{H}$) and heterogeneous models ($\mathcal{E}$) in Table \ref{tbl:results}. We find that the heterogeneous model which incorporates both the user network and domain webgraph outperforms all other models with an average accuracy over 10 runs of $0.7865 \pm .002$. However, swapping the Heterogeneous GraphSage Convolutions in this model with a Heterogeneous Graph Transformer ($HetGT$) convolutions yielded a slightly better F1 ($0.789 \pm .003$). While the accuracy of the Social Media user model ($\mathcal{H}_{users}$ is higher, this model is only evaluated the 2,475 labeled domains in $\mathcal{W}_{target}$ to which social media users were connected, and thus are not directly comparable to other models. In the heterogeneous models, we considered embedded user text features (text) as well as user node node2vec features (n2v). Interestingly, including embedded text of social media users as node performed worse than using features that uniquely consider network position.

\begin{table*}[!ht]
\centering
\begin{tabular}{lll}
\hline
\multicolumn{1}{c}{Model} & \multicolumn{1}{c}{Accuracy} & \multicolumn{1}{c}{F1}  \\
\hline
$\mathcal{H}_{domains}$                 & 0.7686 ± .004                & 0.7545 ± .005           \\
Carragher et al.          & 0.7799 ± .002                & 0.7714 ± .002           \\
$\mathcal{H}_{users}^*$            & 0.8113* ± .007                & 0.7582* ± .009           \\
$\mathcal{H}_{domains+users}$                 & 0.7696 ± .006                & 0.7559 ± .006           \\

$\mathcal{E}_{domains+users(n2v)}$     & \textbf{0.7865 ± .002}       & 0.7777  ± .003 \\
$HetGT_{domains+users(n2v)}$     & 0.7820 ± .003       & \textbf{0.7895  ± .003} \\
$\mathcal{E}_{domains+users(text)}$         & 0.7738 ± .118                & 0.7657 ± .018    \\
$\mathcal{E}_{domains+users+dredge}$          & 0.7787 ± .004                & 0.7675 ± .007    \\ 
\hline
\end{tabular}
\caption{Mean Accuracy, F1, and standard deviations for each GNN ablation over 10 runs. $\mathcal{H}$ denotes homogeneous and $\mathcal{E}$ denotes heterogeneous. * denotes a social-media-only model run and evaluated on the 2,475 labeled domains mentioned in the Twitter data.}
\label{tbl:results}
\end{table*}

We compare our best performing model with the approach proposed in \citet{yang2023large} in Table \ref{tbl:gptcomparison}. The authors in \citet{yang2023large} used ChatGPT to return the probability that 7,318 websites---7,317 of which are in our list---were unreliable labeled domains which we will denote $\mathcal{W}_{LLM}$. Following the authors, we use a cutoff of 0.5 to binarize their ChatGPT predictions. Our model was trained on many of the 7,318 websites that GPT evaluated, so comparing all data results in label leakage. Consequently, we provide a second evaluation, where we only compare labeled websites that our models never encountered in training, i.e.,  $\mathcal{W}_{LLM} \cap W_{test}$. On both $\mathcal{W}_{LLM}$ and the $\mathcal{W}_{LLM} \cap W_{test}$ evaluations, our model outperformed ChatGPT. Our model yielded an F1 that was 0.03 higher than ChatGPT on $\mathcal{W}_{LLM}$ and 0.04 higher on  $\mathcal{W}_{LLM} \cap W_{test}$.

\begin{table}[!h]
\centering
\begin{tabular}{lll}
\hline
\textbf{}                 & \textbf{Acc.} & \textbf{F1}    \\
\hline
GPT ($\mathcal{W}_{LLM}$)  & 0.81              & 0.751          \\
$\mathcal{E}_{domains+users(n2v)}$ ($\mathcal{W}_{LLM}$) & \textbf{0.837}    & \textbf{0.780}  \\
GPT ($\mathcal{W}_{LLM} \cap \mathcal{W}_{test}$)    & 0.782    & 0.701 \\
$\mathcal{E}_{domains+users(n2v)}$ ($\mathcal{W}_{LLM} \cap \mathcal{W}_{test}$)           & \textbf{0.819}    & \textbf{0.745}\\
\hline
\end{tabular}
\caption{Our best model outperforms the LLM approach proposed in \citet{yang2023large}. We evaluate against all 7,318 websites the authors considered ($\mathcal{W}_{LLM}$) and against the subset of $\mathcal{W}_{LLM}$ contained in our test set ($\mathcal{W}_{LLM} \cap \mathcal{W}_{test}$).}
\label{tbl:gptcomparison}
\end{table}

\subsection{Unreliable Domain GNN Discovery}

Two annotators\footnote{25-30 year-old PhD candidates studying misinformation} independently annotated the reliability of the top 20 predictions (sorted by prediction confidence) of the top performing model, heterogeneous model domains+users(n2v), over the set of all unlabeled domains. Inter-annotator agreement in the first round had a Krippendorff's $\alpha = 0.78$. The annotators then met and resolved the single disagreement. We compare our model with the webgraph-based discovery (WG-BD) approach in \citet{carragher2024detection}\footnote{As that paper's discovery process did not rank discovered domains, we sort results by the misinformation classifier proposed in the work.}. Again, two independent annotators ranked the results, yielding a Krippendorff's $\alpha = 0.69$. Additional annotation details are discussed in Appendix D.

For comparison with previous unreliable domain discovery approaches, we additionally report partial F1, defined in \citet{domain_discovery_social_media} as the set of true positive unreliable domains with credibility labeled ``mixed'' or worse by Media Bias Fact Check. Thresholding at a prediction confidence level of 0.7 (see Appendix C for sensitivity analysis), the Partial F1 of our top-performing GNN model is 0.25, which is largely  previous works; 0.29 for $\text{SM-BD}$ \cite{domain_discovery_social_media}) and 0.28 for $\text{WG-BD}$ \cite{carragher2024detection}. As demonstrated in Table \ref{tbl:discovery}, we observe that our heterogeneous domains+users model outperforms both competing systems at each level of precision we consider.


\subsection{Dredge Word Discovery}
We consider two dredge word discovery approaches: one approach ($\dwbd_{lower}$ and $\dwbd_{upper}$) adapted from the discovery approach proposed in \citep{carragher2024detection} and the other using the GNN trained with dredge word context ($\mathcal{E}_{domains+users+dredge}$), which we shorten to ($\mathcal{E}_{d+u(n2v)+dredge}$) in Table \ref{tbl:discovery}. The GNN without dredgewords outperforms all other baselines at every level of precision tested. We note that $\dwbd_{lower}$ yields higher precision than the webgraph based discovery baseline and the $\dwbd_{upper}$ baseline at P@5 and P@10, but the reported social media P@20 outperforms the webgraph and dredge-word discovery processes. Several of the sites returned by $\dwbd_{upper}$ were user-based forums, crypto exchanges, or supplement sellers, which while not necessarily reliable, are out-of-domain and not rated as unreliable.

In the latter approach, as a consequence of 1) only extracting dredge word networks for only 46 of the least reliable domains 2) the dominance of social media and shopping sites in the dredge word SERPs, the dredge word GNN learns to heavily associate social media sites and online shopping platforms with unreliable domains. The 10 most confident predictions the GNN returns contain 7 social media sites, amazon, itunes, and s3.amazonaws. We find at least several of the pseudo-scientific and conspiratorial dredge words are targeted by dubious sellers, podcasters, and social media influencers. As these websites are out-of-domain for our task, but still often contain unreliable content, we chose not to evaluate Precision@k for this set. A deeper exploration of the link between these conspiratorial dredge words and non-news domains would be a fruitful path for future work.

\begin{table}[]
\centering
\addtolength{\tabcolsep}{-0.25em}
\begin{tabular}{lllllll}
\hline
Model     & P@5 & P@10 & P@20 & PF1& PP & PR\\
\hline
$\mathcal{E}_{d+u(n2v)}$       & \textbf{1}   & \textbf{1} & \textbf{0.9} & 0.25  & 0.18 & 0.35 \\
$\mathcal{E}_{d+u(n2v)+dredge}$      & * & * & * & 0.07 & 0.04 & 0.26\\
SM-BD & - & - & 0.7 & \textbf{0.29} & \textbf{0.24} & \textbf{0.37}\\
WG-BD & 0.2 & 0.5  & 0.65 & 0.28 & 0.24 & 0.31 \\
\dwbd$_{lower}$      & 0.8 & 0.7 & 0.55 & 0.02 & 0.17 & 0.01 \\
\dwbd$_{upper}$      & 0.6 & 0.4 & 0.5 & 0.12 & 0.28 & 0.08 \\
\hline
\end{tabular}
\caption{Precision@K and Partial F1 (PF1), Partial Precision (PP), and Partial Recall (PR) discovery evaluations.
\cite{domain_discovery_social_media}.}
\label{tbl:discovery}
\end{table}

\section{Analysis and Discussion}

We explore the 25 most confident unreliable predictions of our best performing model over the set of all unlabeled domains in $\mathcal{W}_{source}$. We observe that these predictions seem to fall into three broad categories: health misinformation (6 websites), Qanon\footnote{A conspiracy theory that claims that a cabal of Satanist, sex-trafficking, child molesters is running the world.} misinformation (5), and German-language far-right misinformation (9). In all three categories, the majority of the websites express skepticism towards vaccines, the existence of COVID-19, or both. Most interestingly, our investigation of the these categories captures an interplay between websites spreading medical misinformation and the websites profiting off of it.

Six of our model's 25 most confident predictions, are an likely co-owned network of websites selling dubious medical products. The $\mathcal{W}_{target}$  website ``holitichealth''\footnote{\url{https://web.archive.org/web/20240329171920/https://holistichealth.one/}} offers users  advice on how to detox from vaccines, which the site claims ``poison your DNA, brain, nervous system and immune system''. The six sites that most frequently link to holistichealth all appeared in our model's most confident unreliable predictions---all sites have the same HTML template, end in .one,  and resolve to the same IP address, and frequently link to one another, heavily suggesting co-ownership. Our model's second most confident prediction,  herbalremedies\footnote{\url{https://web.archive.org/web/20240225051441/https://herbalremedies.one/herbs-for-cancer/}}, sells 17 supplements that the site claims kill cancer cells, including colloidal silver and ''parasite cleansing herbs''. Another .one site, ``emfprotection'', sells electric and magnetic field protection gear that the site claims will defend customers from psychic mind control and 5G radio waves. While we did not observe any Twitter mentions of the 6 unlabeled sites, holitichealth, was mentioned by 7 Twitter accounts. Interestingly, the 7 Twitter accounts often linked to some reliable news sites, but also linked to Zero Hedge, The Epoch Times, Breitbart, and other low-and-mixed reliability websites. 

The second large category of discovered sites promoted conspiratorial---often Qanon \cite{garry2021qanon}---content. Three of these sites (wakeup.icu, vintel1776.net, and qaggregator.news), are now dead, but these sites were all active during the period that Twitter data were collected. Two of the sites linked heavily to a website containing all of Q's drops\footnote{\url{https://web.archive.org/web/20240911060604/https://qagg.news/}}, and wakeup linked heavily to a Qanon discussion board. The only identified Qanon website that was mentioned on Twitter was Hnewswire\footnote{\url{https://web.archive.org/web/20240903154602/https://hnewswire.com/}}, which was mentioned by 100 distinct Twitter accounts that also heavily mentioned Zero Hedge, Gateway Pundit, and many other low and mixed reliability domains. Hnewswire's navigation menu contains a trending topics bar which includes categories like ``Covid Kill Shot'', ``Plandemic'', and ``Demonic Activity End Times''. Independent annotations of predictions can be found in the Github repository.

The top 25 discovered websites were mentioned by 173 Twitter users in the COVID data who cumulatively mentioned over 3,165 domains. Interestingly, 42 of these users also used dredge words at least once in COVID-19-related tweets during that time period. Most of these are names of unreliable websites, e.g., ``creation'', ``gateway pundit'', ``rense'', ``infowars'', and ``bitchute'', but the dredge words ``what really happened'', ``national vaccine'', ``being 6'', and ``missing links'' were also used by this set of users. 

This analysis highlights the connectivity of COVID-19 misinformation. Both the Qanon websites and the far-right German-language websites (which we do not discuss due to space constraints) actively promoted COVID misinformation, often alongside other conspiracies. This misinformation was then monetized by the .one websites, which sell, for example, ``natural Hydroxychloroquine and Ivermectin alternative[s]'' to those concerned about COVID\footnote{\url{https://web.archive.org/web/20240420185505/https://holistichealth.one/natural-hydroxychloroquine-and-ivermectin/}}. This is a key advantage of the current method over article-based approaches; incentive structures are often built into networks. There is often overlap in those spreading conspiracies and those monetizing them \cite{ballard2022conspiracy}, and having a system that considers the diverse paths that users can take to unreliable content helps us identify that interplay.


\section{Limitations}

While we attempted to mitigate limitations where possible, several substantial ones exist. First, the Twitter data were collected from 2020-2022, but the dredge words used in this paper were based on Google SERPs in 2024. We cannot say that the dredge words we extracted surfaced the same content in 2024 as when the phrases were initially used on Twitter. Further analysis on the temporal alignment of keyword rankings and is warranted and would be a promising avenue for future research. 

As with the Ukraine Biolabs example, there are also likely many cases where unreliable websites ranked highly, but were overtaken by fact checkers; our approach would not capture this. Future work could explore the impacts of better temporal alignment, which would have the added benefit of covering data voids formed around breaking news stories. Future work could also attempt to expand this work beyond Google.

A key limitation is that the current process for extracting dredge words relies on paid third-party services. While we validated each of the extracted dredge words to ensure they are ranking in Google's top-10, approaches to extract keyphrases from, for example, common-crawl webgraphs would provide more democratization of this line of research. We publish all dredge word data collected for this project in line with that goal. We additionally note that due to resource and domain expert availability constraints, the discovery process was only evaluated by two domain experts; we release all annotations and annotator notes alongside the code. Additionally, the Partial F1 metric we use for discovery evaluation is limited, as previous work has shown many PoliticalNews and MBFC sites on which the metric relies are dead or no longer active \cite{carragher2024detection}. However, to our knowledge, no better metric exists for evaluating unreliable domain discovery systems. This is a needed avenue for future work. 

Finally, dredge words were only extracted for 46 of the least reliable 11,327 labeled domains. Their inclusion in our models is therefore not particularly helpful. We elected to include the results from these models because they point to an interesting and understudied phenomenon: bidirectional paths between social media and search engines, which appear to be often targeted by monetary interests. In future work, we plan to repeat this process with dredge words extracted for a larger set of unreliable domains, along with keywords extracted from reliable domains.
\section{Conclusion}

Proactive content moderation requires systems that rapidly and continuously identify unreliable information sources. By considering additional direct paths that users can take to unreliable websites, we can improve the ability of models to identify these sources. We propose the idea of \textit{Dredge Words} and highlight their bidirectional connections with social media. Finally, we demonstrate that our best-performing model outperforms competing systems on the tasks of domain reliability classification and unreliable domain discovery. This work demonstrates the signals present in the direct and indirect pathways that users follow to unreliable websites. Better understanding these paths could be a promising area of collaboration between platforms, organizations, and researchers intent on mitigating the spread of misinformation across digital ecosystems.

\section{Acknowledgements}
The research for this paper was supported in part by the Office of Naval Research, Minerva-Multi-Level Models of Covert Online Information Campaigns, Office of Naval Research under grant N000142112765, ARMY Scalable Technologies for Social Cybersecurity under grant W911NF20D0002, and by the Knight Foundation. It was also supported by the Informed Democracy and Social-cybersecurity Institute (IDeaS) and the center for Computational Analysis of Social and Organizational Systems (CASOS) at Carnegie Mellon University. The views and conclusions are those of the authors and should not be interpreted as representing the official policies, either expressed or implied, of the ONR or the US Government.

\bibliography{dredgefinal}
\section{Ethical Statement}

To avoid unintentionally boosting the search rankings of unreliable domains, we do not directly link to any referenced unreliable domains. For unreliable sites, we link to either archive.org or third-party reliability assessments of those domains. We further acknowledge the inherent challenges in determining the ``reliability'' of websites, and note that even reliable websites can occasionally publish inaccurate content. We defaulted entirely to third parties in our original acquisition of labels, but emphasize that any reliability-labeling ontology should be scrutinized to minimize harms. We would argue that the system we proposed could be generalized to any reasonable reliability ontology. 

Acknowledging that the Twitter data likely contains examples of ``sensitive events'' which could compromise the privacy or safety of users, we adhere to Twitter's terms of service and do not release this data\footnote{https://developer.x.com/en/developer-terms/agreement-and-policy}. Finally, we acknowledge that by publicly releasing these dredge-word lists, bad actors could use these terms to direct people to unreliable information. However, once data voids are identified, they are often plugged \cite{norocel2023google}. We hope that by releasing this data, reliable websites can target these phrases to reduce both the visibility and the ease of access of unreliable content.

There are a number of ways that any system built to limit misinformation could be abused. These include using the methodology for targeted censorship, the risks of false positives and false negatives, or abusing the system to promote or demote content based on attributes other than reliability. While these risks are real, misinformation is a pervasive global problem. We believe the benefits of advancing our understanding of how misinformation spreads between social media and search engines outweigh these risks.

\section{Acknowledgements}
The research for this paper was supported in part by the Office of Naval Research, Minerva-Multi-Level Models of Covert Online Information Campaigns, Office of Naval Research under grant N000142112765, ARMY Scalable Technologies for Social Cybersecurity under grant W911NF20D0002, and by the Knight Foundation. It was also supported by the Informed Democracy and Social-cybersecurity Institute (IDeaS) and the center for Computational Analysis of Social and Organizational Systems (CASOS) at Carnegie Mellon University. The views and conclusions are those of the authors and should not be interpreted as representing the official policies, either expressed or implied, of the ONR or the US Government.

\appendix
\section{Appendix A: Additional Data Details}
\label{app:data}

 To our knowledge, the list created in \cite{lin2023high} is the most comprehensive public domain reliability rating list as of August 2024. The domains in this list are ranked by the calculated first principal component where low scores correspond to high expert agreement on unreliability and high scores correspond to expert agreement on reliability. While these scores are very sensible across wide principal component score gaps, locally the relative orderings are less clear. Treating website reliability as continuous also muddies the interpretability and the discovery process, as there's not a transparent reason a website should have a score of 0.571 as opposed to 0.570. To address this issue, we elected to binarize the website reliability labels. After manual inspection of the data, we elected to consider the bottom two quintiles as unreliable (corresponding to a principal component threshold of 0.5162). 

From Ahrefs, we were unable to pull backlinks for 193 of the domains. A random sample of 10 of 193 domains and found 9 of them were either dead or unreachable. We therefore elected to drop these 193 domains, leaving us with 11,327 unique reliability-labeled domains and 32,431 unique unlabeled backlinking domains for a total of 43,758 domains. For each of these 43,758 domains, we pull 23 attributes, which for an individual domain, contains fields like the total number of backlinks and outlinks, number of backlinks coming from .edu or .gov domains, and number of referring pages \footnote{\url{https://ahrefs.com/api/documentation/metrics}}.

\begin{figure}[!htbp]
    \centering
    \includegraphics[height=5.5cm]{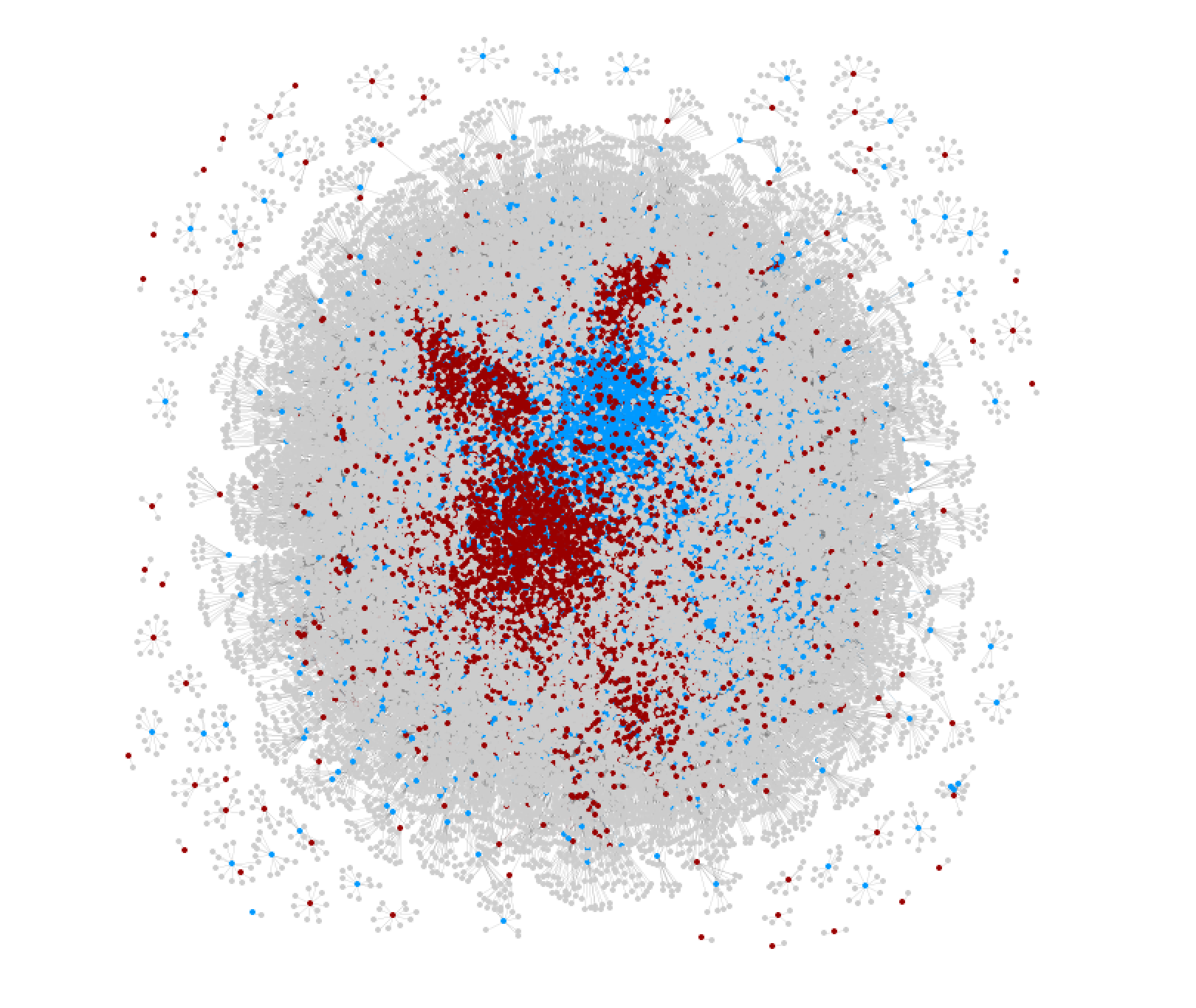}
    \caption{Webgraph colored by domain reliability labels. The network contains 6,861 reliable (blue) websites, 4,466 (red) unreliable websites, and 32,431 unlabeled (grey) backlinking websites.}
    \label{fig:network_viz}
\end{figure}

Similar to the backlink network constructed in \cite{carragher2024detection}, there is clear assortativity across label-reliability groups as can be seen in Figure \ref{fig:network_viz}. The prevalence of unlabeled nodes in our network overwhelms a node attribute assortativity coefficient calculation $r_A = -0.336$. We re-calculate node attribute assortativity on an induced subgraph that only maintains edges between labeled nodes. The subgraph contains less structure and 176 components, but nevertheless displays strong positive associativity $r_A = 0.376$. In other words, domains of the same (binary) reliability are more likely to link to one another than linking to a domain of a different reliability. This provides a strong justification for the use of network-based models.

All GPU experiments experiments were run on a single NVIDIA GeForce RTX 3080. Given the small number of labels, no model took over 10 minutes to train. GNN experiments were implemented in Python3.10 using Pytorch Geometric \cite{Fey/Lenssen/2019}. We publicly release all code, annotations, and non-Twitter data.

\subsection{Twitter Cleaning}
\label{app:tclean}

All twitter data comes from the dataset of 3.6 billion tweets described in the Data section. We first extract all tweets that link to or mention one of our 11,327 websites and all tweets that retweet, reply to, or quote a tweet containing one of those 11,327 websites. Once we have this data, we drop all users that do not explicitly include a link to one of our 11,327 target domains. Second, users with fewer than 10 observed tweets over the time period were dropped. To further condense the dataset, we only include tweets which appeared at least 3 times in the dataset---e.g., if a tweet was reposted or retweeted twice. As reposting and retweeting are important influence metrics on Twitter, this drops the tweets that likely did not receive as much attention. We also dropped domains that were only linked to by a single user and we dropped users that only linked to a single domain, as we hypothesized these pendulum nodes would be of limited use given the size of the graph. As a result of the expense of the webgraph attribute API, we chose to further restrict the Twitter data to only include tweets that mention at least one of the 43,758 domains for which we extracted attributes. This corresponds to keeping tweets that mention at least one of the 11,327 labeled domains for which we have attributes or tweets that mention one of the 11,327 and co-mention any of the 32,431 unlabeled domains.


\subsection{Dredge EDA}
\label{app:dredge_eda}

We applied the python library langdetect \cite{Danilak2014} to the list of 3,933 dredge words used in the paper, but we observe that automated language extraction of dredge words is a challenge. We observed many dredge words are disembodied fragments, names of people, or simply foreign words adopted by various communities. For example, several Hindi-language-origin yoga poses appear in the dredge word list: parivrtta anjaneyasana and parivrtta utkatasana. These terms surface \textit{gaia}, a conspiracy pseudoscience website\footnote{\url{https://mediabiasfactcheck.com/gaia/}}. There are additionally arabic and chinese dredge words that surface \textit{creation}, a website that frequently promotes pseudoscience\footnote{https://mediabiasfactcheck.com/christian-ministries-international/}. Qualitatively, hatespeech is fairly uncommon in the dredge words we identified; 11 unique dredge word phrases contain the non-euphemistic writing of the n-word and 3 others contain a homophobic slur.



\section{Appendix B: Curriculum EDA}
\label{app:curr}
 Among the websites rated most reliable, the style, content, and widespread name recognition would very likely result in most annotators to correctly identifying the sites as reliable (e.g., reuters.com, nasa.gov, smithsonianmag.com, nature.com). The landing pages of some of the least reliable labeled domains are also typically easy to categorize, with websites that contain toxic green backgrounds accompanied by blurry photos of George Soros \footnote{\url{https://web.archive.org/web/20240112053008/http://www.endgamethemovie.com/}}, proclamations of white nationalism \footnote{\url{https://web.archive.org/web/20240108044820/https://www.stormfront.org/forum/}}, or full links to alien siting websites and paid psychic services \footnote{\url{https://web.archive.org/web/20240116140224/https://rense.com/}}. Though our webgraph model is content-agnostic, we hypothesize that the webgraph and site attributes will exhibit similarly "easy" patterns across the extremes. To test this hypothesis, we implement a curriculum learning batching procedure \cite{bengio2009curriculum}.

\section{Appendix C: Partial F1 threshhold sensitivity}
\label{app:pf1}
\begin{figure}
    \centering
    \includegraphics[width=\linewidth]{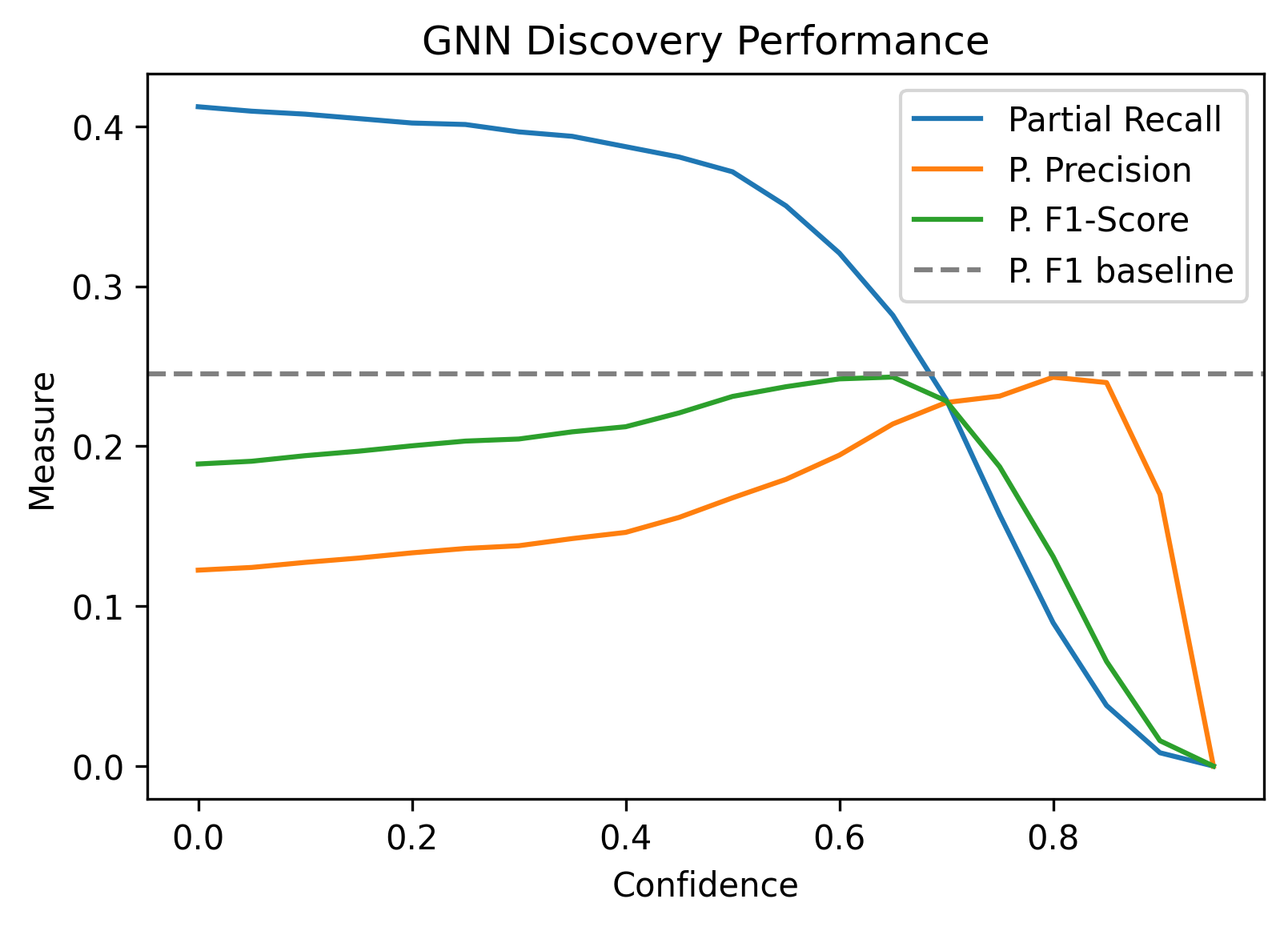}
    \caption{GNN discovery performance vs. classifier confidence reveals that the Partial F1 metric is precision bounded.}
    \label{fig:gnn_discovery_confidence}
\end{figure}

To investigate the partial F1 metric we experiment with various classifier confidence thresholds. Naturally, as confidence increases, we see higher precision and lower recall (Figure \ref{fig:gnn_discovery_confidence}). The disparity between partial recall and partial precision implies that our discovery process is not very precise. However, in the manual evaluation of Precision@k, we find precision is actually very high. This indicates a fundamental issue with the partial precision metric; it underestimates model performance, particularly as the MBFC reference list ages. This is in line with previous findings that unreliable domain lists quickly become outdated \citep{carragher2024detection}, and it complicates the evaluation of discovery processes. Further work on domain reliability discovery metrics is needed.

For Precision@k assessments, two annotators were asked ``is this website unreliable'' and asked to provide a rationale for why or why not. Annotations involved exploration of the domain and lateral reading---looking at what other reliable sources have written about the domain. We publicly release these annotations alongside annotator rationales.\footnote{Following anonymity period.}

\section{Appendix D: Annotation details}

Annotators were asked to assess on a binary scale, whether each identified site was reliable or unreliable. In some of our discovered sites, this distinction is unambiguous (e.g. prominently pro-Q-anon blogs). If the website appeared to express political opinions, the annotators were instructed to search for medical or scientific claims. If authors made conspiratorial (e.g., ``covid is part of a great reset'') or pseudo-scientific claims (e.g., ``global warming is a hoax''), the website was judged to be unreliable. Annotations were done independently, and we report Krippendorf's Alpha on the independently-annotated lists. In both annotation sets, annotator agreement on website reliability was substantial with $\alpha = 0.78$ and $\alpha = 0.69$. While we do not control for potential annotator bias, annotators never used political bias as a rationale for rating a site as unreliable. We release all annotations along with notes left by the annotators.

\end{document}